\documentclass[11pt]{article}

\usepackage[margin=1.05in]{geometry}
\usepackage[T1]{fontenc}
\usepackage{lmodern}
\usepackage{booktabs}
\usepackage{graphicx}
\usepackage{xcolor}
\usepackage{float}
\usepackage{pgfplots}
\pgfplotsset{compat=1.17}
\usepackage{tikz}
\usetikzlibrary{arrows.meta,positioning,shapes.geometric}
\usepackage[hidelinks]{hyperref}

\definecolor{mercblue}{RGB}{27,111,196}
\definecolor{blazegray}{RGB}{110,116,125}

\title{The Atoms of the Score:\\
Record-Level versus In-Engine Composite Evaluation\\
of Clinical Quality Language}
\author{Angelo Kastroulis\\Carrera Group\\\texttt{akastroulis@alumni.harvard.edu}}
\date{Draft v2 --- July 2026}

\begin{document}
\maketitle

\begin{abstract}
Clinical Quality Language (CQL) engines serve two axes of clinical computation: decision
support --- evaluate one patient, now --- and quality measurement --- score a population
against a measure. Quality measurement is itself multidimensional: the composite score is
rarely the end product, because the individual determinations that make it up are what allow
a score to be inspected, attributed, and acted on. Two engine architectures follow from where
that aggregation happens. One computes the composite \emph{inside} the engine over the whole
store --- no data movement, maximal aggregate throughput. The other evaluates individual
records and lets the composite be totaled externally, keeping every intermediate determination
available. Mercury is a purpose-built CQL database engine of the second kind: it treats CQL
evaluation as a database problem --- FHIR resources stored in a compact binary encoding keyed
patient-first, indexes derived from what CQL retrieves filter on, a planner selecting access
paths, and CQL itself as the query language --- rather than as in-memory interpretation over
a generic FHIR store. We evaluate Mercury 2.0.1 against Blaze 1.10.1 --- an engine of the
first kind and among the fastest CQL evaluators available, which is precisely why it is the
right benchmark --- using Blaze's own published suite on identical AWS hardware over a
100{,}000-patient Synthea corpus (112.3\,M resources). The engines disagree by orders of
magnitude in \emph{both directions}: Mercury answers record-level queries 7.5--25.7$\times$
faster at the median (1.0--9.7\,ms vs.\ 26.8--72.6\,ms) and sustains 7.3--9.1$\times$ higher
extraction throughput, while Blaze computes in-engine composites 13--1{,}082$\times$ faster.
Correctness is held to a record-level standard, not merely count equality: a 10{,}000-pair
per-patient agreement gate across engines reached 100.000\,\% after uncovering --- and driving
fixes for --- two ingest bugs that composite-count comparison provably could not detect,
because the aggregates were accidentally correct while individual answers were wrong. Ingest
also favors the purpose-built path (74.7\,min vs.\ 4\,h\,36\,m to query-ready). We attribute
both regimes to a single architectural choice, quantify a ``store hygiene tax'' in which a
correctness fix sped up every lookup 1.2--2.8$\times$, and argue that record-level agreement
gates should be standard practice in engine benchmarking.
\end{abstract}

\section{Introduction}\label{sec:intro}
CQL execution in production systems runs along two axes. The first is \emph{decision
support}: a CDS Hooks call, a point-of-care gap check, a patient-facing application --- one
subject, a millisecond budget, an answer that must arrive while someone is waiting. The
second is \emph{quality measurement}: scoring a population against a measure for reporting,
improvement, or payment. The second axis is more layered than it first appears. The composite
score --- a numerator over a denominator --- is rarely what an organization actually needs;
the \emph{atoms} of the score, the per-patient determinations, are what make the score
explainable and actionable. A rate without its members cannot tell you whom to call, what
documentation is missing, or why the number moved.

Where the aggregation happens therefore divides engine architecture. One family computes the
composite \emph{in the engine}, scanning the whole store: no data movement, one compiled
plan, maximal aggregate throughput --- and the intermediate determinations are a byproduct
the caller may or may not be able to afford to extract. The other family evaluates
\emph{individual records} and leaves totaling to the caller: every atom is first-class and
inspectable, decision support and measurement share one execution path, and the composite is
a trivial external sum --- at the cost of paying per-record evaluation overhead as many times
as there are records. Neither choice is free. This paper measures both sides of that trade
with one corpus on one machine class.

This paper also takes a position: \emph{an atom-level methodology is more useful than an
aggregate totaling machine}. Three observations carry the argument. First, the atoms serve
both axes --- the same record-level evaluation that answers a decision-support call is the
path to a quality measure totaled externally --- while a composite-only result serves one.
Second, the information flows one way: the composite is recoverable from the atoms by
trivial summation, but the atoms are not recoverable from the composite. Third, and most
consequentially for benchmarking, correctness is only \emph{checkable} at the atom level:
Section~\ref{sec:journey} presents two bugs that left aggregate counts exactly correct while
individual answers were wrong --- a failure mode no amount of composite validation can
detect. The price of this position is measured rather than hidden: when the composite truly
is all that is needed, the in-engine scan is dramatically faster
(Section~\ref{sec:regimeb}), and we report that with the same prominence as our wins.

\paragraph{Why a purpose-built CQL database engine.} The prevailing reference stacks do not
translate CQL to SQL. They interpret it --- compiled to ELM --- in memory, typically on the
JVM, against FHIR resources fetched from a generic FHIR server. What they do not do is treat
evaluation the way a database would: store the data compactly in a form chosen to optimize
retrieval, then build the machinery needed to locate and compute --- indexes over the
properties retrieves actually filter on, a planner selecting access paths, an execution
layer that owns its storage. The engine is a calculator bolted to a store that knows nothing
about it; every retrieve is a round-trip into somebody else's data layout, parsed from JSON
on arrival. Building the engine \emph{as} a database instead lets CQL evaluation inherit
what decades of data-systems research made routine --- compact binary encodings and
compression, zero-copy reads, indexing, access-path selection, query planning --- with the
healthcare-specific nuances baked in: the FHIR resource as the native record, the patient
partition as the unit of physical locality (matching CQL's unit of evaluation, the subject
context), and index keys derived from CQL retrieve shapes. Blaze took this database turn for
the population axis --- it owns its store and compiles queries against it, which is why it is
fast. Mercury takes the same turn for the record axis, and is deliberately an engine of the
record-level family. That makes Blaze the appropriate benchmark: if the record-level design
cannot hold its own against the best database-shaped engine of the other family, the design
is not interesting.

This paper makes five contributions. (1)~A two-regime benchmark design --- interactive
record-level latency, concurrent record-level extraction, and in-engine composite
aggregation --- layered on Blaze's own published query suite and protocol: the defender's
home field. (2)~Head-to-head results on identical hardware and corpus in which each engine
wins its regime by one to three orders of magnitude; both results are true simultaneously
and follow from one storage decision. (3)~A record-level cross-engine agreement gate
(10{,}000 paired booleans) as a correctness methodology, with case studies in which it caught
bugs invisible to aggregate-count equality --- including one where the aggregate was
\emph{accidentally correct}. (4)~A complete failure taxonomy: both engines fail the same
terminology-dependent queries loudly; one disclosed DNF; zero silent wrongs after fixes.
(5)~Ingest-path attribution isolating allocator policy ($\sim$8$\times$), lock removal
($\sim$4$\times$ parallel-phase, 1.3$\times$ end-to-end), and a merge-operator experiment
($\approx$ nil).

\section{Background}\label{sec:background}
\paragraph{CQL and measure evaluation.} A FHIR \texttt{Library} carries CQL source; a
\texttt{Measure} references it; \texttt{\$evaluate-measure} evaluates an initial-population
expression per subject (\texttt{reportType=subject}) or over the store
(\texttt{reportType=population}). The two report types are the API surface of the two axes
above.

\paragraph{The CQF operation surface is wider than \texttt{\$evaluate-measure}.} The CQF
framework defines a family of record-level operations beyond measure evaluation:
\texttt{Library/\$evaluate} (evaluate any named expression for a subject), the system-level
\texttt{\$cql} operation (ad-hoc expressions), \texttt{\$data-requirements}, and
\texttt{\$package}. These are where CQL is used as a \emph{language} --- the substrate of
decision support and application logic --- rather than as a measure-report generator, and a
benchmark centered on CQL arguably ought to make them its focus: \texttt{\$evaluate-measure}
is one case, and for record-level workloads a marginal one. Blaze's published suite, which we
adopt verbatim as the defender's home field, exercises \texttt{\$evaluate-measure} only; a
dedicated cross-engine benchmark of the full CQF record-level surface (\emph{CQF-Bench}) is
in development by the authors and is out of scope here.

\paragraph{Blaze.} A Clojure FHIR server that compiles CQL once per evaluation and iterates
the patient axis in a tight scan; amortized per-patient cost on that axis is sub-microsecond.
Blaze publishes its benchmark suite (queries, Synthea recipe, timing protocol, an
\texttt{eval-duration} extension) in its repository; we adopt all of it.

\paragraph{Mercury.} A Rust engine over RocksDB, version 2.0.1 evaluated here. Every
clinical row is keyed \texttt{patient|type|id} with index postings
\texttt{patient|property|value}; the store is physically partitioned by subject. Evaluation
binds one subject's context and reads only that subject's keys. Around that storage core
Mercury carries the full database toolchain: a query planner with access-path selection
(index routes vs.\ scans, chosen per retrieve), a compact FHIR-shaped binary record format
with zero-copy reads, intelligent valueset-and-code matching (terminology resolved to direct
code filters at compile time, unresolvable references failing loudly), and the CQF operation
surface (\texttt{\$evaluate-measure}, \texttt{Library/\$evaluate}, \texttt{\$cql},
\texttt{\$data-requirements}, \texttt{\$package}). The design intent is the record-level
family: O(1) subject workflows, simple deletion and compaction, atoms as the primary product;
in-engine composite scans were the known un-optimized path.

\section{Method}\label{sec:method}
\subsection{Test setup: corpus, hardware, physical footprint, load time}

\begin{table}[H]
\centering\footnotesize
\caption{Test setup. One engine resident at a time on the same instance class; stores
snapshotted to object storage for reproduction. Ingest paths differ by design and are
declared: Blaze ingests FHIR transaction bundles over HTTP (blazectl, 16-way concurrent);
Mercury ingests NDJSON through its bulk path --- the same resources, reference-rewritten per
the FHIR transaction rules.}
\label{tab:setup}
\begin{tabular}{@{}l p{0.60\linewidth}@{}}
\toprule
Corpus & Synthea (seed 3256262546): 100{,}000 patients, \textbf{112{,}322{,}471 resources}\\
Raw corpus & 100{,}000 gzipped transaction bundles, \textbf{17.8\,GB}\\
NDJSON (Mercury path) & 20 gzipped shards, \textbf{10.4\,GB}\\
Hardware & AWS \textbf{r6i.8xlarge}: 32 vCPU, 256\,GB RAM; gp3 1\,TB, 16\,K IOPS\\
Blaze & 1.10.1, documented benchmark config: 64\,GB heap, 64\,GB block cache\\
Mercury & 2.0.1, community engine: 32\,GB block cache, 30 eval threads\\
Store on disk & Blaze \textbf{127.3\,GB}; Mercury \textbf{185.0\,GB} (binary FHIR + patient-scoped postings)\\
Load time & Blaze \textbf{4\,h\,36\,m} ($\approx$6{,}800 res/s); Mercury \textbf{74.7\,min} ($\approx$25{,}100 res/s) --- \textbf{3.7$\times$}\\
Protocol & Blaze's 9-run protocol, first two discarded; container restart between queries\\
\bottomrule
\end{tabular}
\end{table}

\begin{table}[H]
\centering\footnotesize
\caption{Resource-type profile (exact counts). Observation-dominated --- Synthea's
longitudinal-record shape --- averaging $\sim$1{,}123 resources ($\sim$640 observations) per
patient. Consistency check: the Condition row count equals the
\texttt{stratifier-condition-code} strata total both engines reported
(Table~\ref{tab:population}).}
\label{tab:profile}
\begin{tabular}{lrr}
\toprule
Resource type & Count & Share\\
\midrule
Observation              & 63{,}950{,}790 & 56.9\,\%\\
DiagnosticReport         & 17{,}237{,}029 & 15.3\,\%\\
Procedure                & 15{,}615{,}811 & 13.9\,\%\\
Encounter                &  9{,}551{,}864 &  8.5\,\%\\
Condition                &  5{,}658{,}095 &  5.0\,\%\\
MedicationAdministration &     104{,}441  &  0.1\,\%\\
Medication               &     104{,}441  &  0.1\,\%\\
Patient                  &     100{,}000  &  0.1\,\%\\
\midrule
\textbf{Total}           & \textbf{112{,}322{,}471} & 100\,\%\\
\bottomrule
\end{tabular}
\end{table}

Tables~\ref{tab:setup} and~\ref{tab:profile} summarize the setup and corpus. Declared
deviations: Mercury lacks the FHIR transaction interaction for artifact creation (individual
POSTs); Mercury received \texttt{reportType=population} explicitly; and unscoped composite
evaluation --- disabled and reported \emph{not supported} in Mercury as shipped --- was
explicitly enabled for the Regime-B runs.

\subsection{Before the test: the ingest observation}\label{sec:ingestobs}
The timed protocol begins at a loaded store, so ingest is not part of the benchmark --- but
getting both engines to that starting line was itself instructive, and we report it as an
observation. Blaze ingested the corpus the way its documentation prescribes: 16-way
concurrent FHIR transaction bundles over HTTP, sustaining $\approx$6{,}800 resources/s for
\textbf{4\,h\,36\,m} end-to-end. Mercury ingested the same resources through its bulk path at
$\approx$25{,}100 resources/s --- \textbf{74.7 minutes} to query-ready, \textbf{3.7$\times$}
sooner on wall clock, including the FHIR-transaction-rule reference rewriting. The paths
differ by design and the comparison is observational, not protocolized; but the operational
difference is real for anyone who must stand up, refresh, or replay a store. Mercury's
ingest surface also extends beyond what this exercise used: bulk FHIR \texttt{\$import},
asynchronous ingest jobs with progress reporting, and direct NDJSON file reads that bypass
HTTP entirely --- the modalities a database offers because loading is part of its job, not an
inconvenience in front of it. The trade is disclosed in the other direction too: Mercury
\emph{pays more disk} for what it builds (185.0 vs.\ 127.3\,GB) --- the patient-first key
space plus per-property index postings are the physical price of the millisecond read path
measured in Section~\ref{sec:regimea}.

\subsection{The two regimes}
\textbf{Regime A (record-level).} A seeded-random sample of 1{,}000 patients (seed 42; sample
SHA-256 published). \emph{A1 interactive}: sequential per-subject
\texttt{\$evaluate-measure}, three passes, first discarded; wall p50/p95/p99/max and
server-side evaluation time via the \texttt{eval-duration} extension (implemented in Mercury
for parity). \emph{A2 extraction}: the same subjects fanned out at concurrency 16 --- the
compute-the-atoms-and-total-externally path to a quality measure. \emph{Full fan-out}: all
100{,}000 subjects at c=16 for one query; the externally-totaled composite this produces is
cross-checked against both engines' in-engine composites. \textbf{Regime B (in-engine
composite).} Blaze's own regime: unscoped evaluation over the store, 9-run protocol.

\subsection{The agreement gate}
Count equality is necessary but not sufficient --- it validates the composite, not the atoms.
For every Regime-A query we record each engine's per-patient initial-population boolean and
join across engines (10 queries $\times$ 1{,}000 patients). Cross-engine subject identity is
itself non-trivial --- Blaze assigns new server ids at transaction ingest --- so the join key
is the Synthea identifier (map published). Section~\ref{sec:journey} shows this gate catching
two bug classes that composite comparison could not.

\section{Results: Regime A (record-level)}\label{sec:regimea}

Mercury is 7.5--25.7$\times$ faster at the median and 3.2--12.1$\times$ at p99
(Table~\ref{tab:regimea}); on eight of ten queries Mercury's p99 ($\approx$2.4--2.7\,ms) sits
an order of magnitude below Blaze's \emph{median}. The worst single Mercury call in
$\sim$20{,}000 timed calls was 59\,ms --- roughly Blaze's median for \texttt{condition-all}.
Server-side, Mercury's typical evaluation is $\approx$0.65\,ms (the remaining $\sim$0.9\,ms
of wall time is HTTP/JSON envelope).

\begin{table}[H]
\centering\footnotesize
\caption{Regime A on the 100\,k store: 1{,}000-patient sample, three passes (first
discarded), concurrency 16 for throughput. Agreement is per-patient boolean equality across
engines.}
\label{tab:regimea}
\begin{tabular}{l rr r rr r}
\toprule
& \multicolumn{3}{c}{A1 p50 (ms)} & \multicolumn{3}{c}{A2 (patients/s)}\\
\cmidrule(lr){2-4}\cmidrule(lr){5-7}
Query & Mercury & Blaze & Speedup & Mercury & Blaze & Speedup\\
\midrule
condition-450-rare      & 9.72 & 72.63 & 7.5$\times$  & 986   & 132 & 7.5$\times$\\
condition-all           & 6.26 & 52.45 & 8.4$\times$  & 1{,}618 & 178 & 9.1$\times$\\
condition-ten-frequent  & 1.57 & 27.91 & 17.8$\times$ & 2{,}397 & 305 & 7.9$\times$\\
condition-ten-rare      & 1.23 & 27.82 & 22.6$\times$ & 2{,}442 & 306 & 8.0$\times$\\
condition-two           & 1.07 & 27.13 & 25.3$\times$ & 2{,}506 & 314 & 8.0$\times$\\
observation-17861-6     & 1.04 & 26.86 & 25.7$\times$ & 2{,}523 & 326 & 7.7$\times$\\
observation-44261-6     & 1.08 & 26.75 & 24.9$\times$ & 2{,}528 & 318 & 8.0$\times$\\
observation-72514-3     & 1.57 & 26.81 & 17.0$\times$ & 2{,}306 & 315 & 7.3$\times$\\
observation-788-0       & 1.06 & 26.81 & 25.3$\times$ & 2{,}531 & 312 & 8.1$\times$\\
observation-8310-5      & 1.17 & 26.80 & 22.8$\times$ & 2{,}463 & 321 & 7.7$\times$\\
\midrule
\multicolumn{7}{@{}l}{Agreement gate: \textbf{10{,}000/10{,}000 = 100.000\,\%} (every query 1{,}000/1{,}000).}\\
\multicolumn{7}{@{}l}{Full 100\,k fan-out (o-17861, c=16): Mercury \textbf{41.0\,s} vs.\ Blaze \textbf{313.2\,s} (7.6$\times$); both 2{,}515 true, 0 errors.}\\
\bottomrule
\end{tabular}
\end{table}

\subsection{The structural finding: a latency floor vs.\ proportional cost}
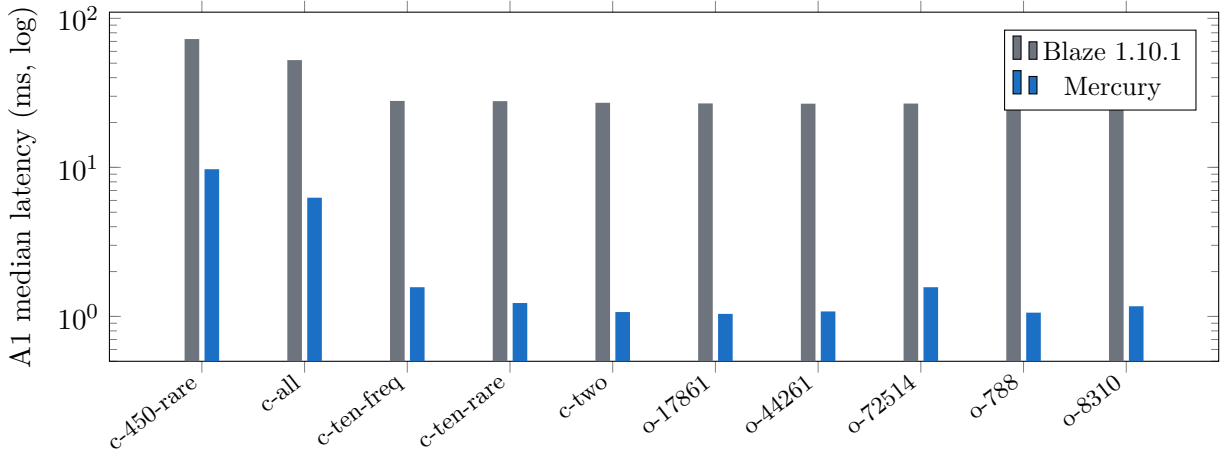
\begin{figure}[H]
\centering
\begin{tikzpicture}
\begin{axis}[
    ybar, ymode=log, log origin=infty,
    width=\linewidth, height=6.2cm,
    ylabel={A1 median latency (ms, log)},
    symbolic x coords={c-450-rare,c-all,c-ten-freq,c-ten-rare,c-two,o-17861,o-44261,o-72514,o-788,o-8310},
    xtick=data, x tick label style={rotate=38,anchor=east,font=\footnotesize},
    ymin=0.5, ymax=110,
    bar width=5.5pt,
    legend style={at={(0.98,0.95)},anchor=north east,font=\small},
    every axis plot/.append style={draw=none},
]
\addplot[fill=blazegray, draw=none] coordinates {
 (c-450-rare,72.63) (c-all,52.45) (c-ten-freq,27.91) (c-ten-rare,27.82) (c-two,27.13)
 (o-17861,26.86) (o-44261,26.75) (o-72514,26.81) (o-788,26.81) (o-8310,26.80)};
\addplot[fill=mercblue, draw=none] coordinates {
 (c-450-rare,9.72) (c-all,6.26) (c-ten-freq,1.57) (c-ten-rare,1.23) (c-two,1.07)
 (o-17861,1.04) (o-44261,1.08) (o-72514,1.57) (o-788,1.06) (o-8310,1.17)};
\legend{Blaze 1.10.1, Mercury}
\end{axis}
\end{tikzpicture}
\caption{Median single-subject latency by query. Blaze pays a $\sim$27\,ms per-evaluation
floor --- eight of ten medians land within 1\,ms of each other regardless of selectivity ---
rising only when the retrieve itself is heavy. Mercury has no comparable floor; its cost is
proportional to data touched. The same architecture inverts in the composite regime
(Table~\ref{tab:population}).}
\label{fig:floor}
\end{figure}

Blaze's median is $\sim$27\,ms on eight of ten queries --- within 1\,ms of each other
regardless of selectivity or result size: a fixed per-evaluation orchestration floor, rising
only when the retrieve itself is heavy. Mercury has no comparable floor; its latency tracks
data touched (1.04--9.72\,ms). Figure~\ref{fig:floor} is that finding in one chart: a flat
line crossing a proportional one.

\begin{figure}[H]
\centering
\begin{tikzpicture}
\begin{axis}[
    width=\linewidth, height=6.4cm, ymode=log,
    ylabel={latency (ms, log)}, xlabel={query (ordered as Table~\ref{tab:regimea})},
    xtick={1,...,10},
    xticklabels={c450,call,c10f,c10r,c2,o17861,o44261,o72514,o788,o8310},
    x tick label style={rotate=38,anchor=east,font=\footnotesize},
    ymin=0.5, ymax=110,
    legend columns=2, legend style={at={(0.5,1.02)},anchor=south,font=\footnotesize,draw=none},
]
\addplot[blazegray,thick,mark=*,mark size=1.6pt] coordinates
 {(1,72.63)(2,52.45)(3,27.91)(4,27.82)(5,27.13)(6,26.86)(7,26.75)(8,26.81)(9,26.81)(10,26.80)};
\addplot[blazegray,dashed,mark=o,mark size=1.4pt] coordinates
 {(1,77.44)(2,55.47)(3,30.19)(4,29.82)(5,29.25)(6,29.06)(7,28.81)(8,29.05)(9,28.76)(10,28.83)};
\addplot[mercblue,thick,mark=*,mark size=1.6pt] coordinates
 {(1,9.72)(2,6.26)(3,1.57)(4,1.23)(5,1.07)(6,1.04)(7,1.08)(8,1.57)(9,1.06)(10,1.17)};
\addplot[mercblue,dashed,mark=o,mark size=1.4pt] coordinates
 {(1,17.86)(2,13.41)(3,3.91)(4,2.72)(5,2.55)(6,2.43)(7,2.70)(8,9.13)(9,2.38)(10,2.69)};
\legend{Blaze p50, Blaze p99, Mercury p50, Mercury p99}
\end{axis}
\end{tikzpicture}
\caption{The tail ladder: p50 (solid) and p99 (dashed) per engine. The bands never overlap ---
on eight of ten queries Mercury's p99 sits below Blaze's p50 by roughly an order of magnitude.
The one visible Mercury tail (o-72514, p99 9.1\,ms) is a single 59\,ms call among 2{,}000.}
\label{fig:tails}
\end{figure}
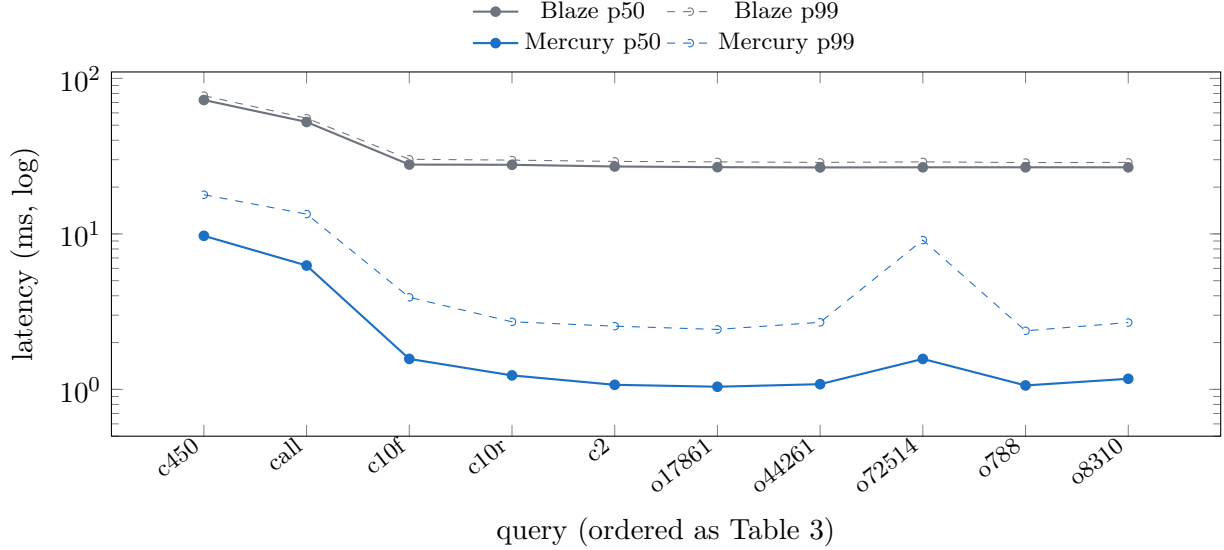

Figure~\ref{fig:tails} extends the finding to the tails: the two engines' latency bands
(p50--p99) do not overlap on any query.

\section{Results: Regime B (in-engine composite)}\label{sec:regimeb}

Blaze is 13--1{,}082$\times$ faster (Table~\ref{tab:population}) --- this regime is what its
architecture is for, and the numbers show it. Mercury matches its composite counts exactly on
all seventeen countable queries, including the 5.66\,M-strata stratifier. One asymmetry
deserves emphasis: unscoped in-engine composite evaluation is not a supported mode of Mercury
as shipped --- the operation is disabled and reports \emph{not supported}; we enabled it
explicitly for this comparison so the regime could be measured rather than merely declared
out of scope. We did not attempt to tune it (Section~\ref{sec:future}).

\begin{table}[H]
\centering\footnotesize
\caption{Regime B: unscoped in-engine composite evaluation, Blaze's 9-run protocol. Every
countable query matches exactly. The three valueset (\texttt{-vs}) queries fail loudly on
both engines (no terminology service in either documented configuration). One disclosed
Mercury DNF: \texttt{stratifier-observation-laboratory-code} (41.65\,M strata).}
\label{tab:population}
\begin{tabular}{l rr rr r}
\toprule
Query & Blaze $n$ & Mercury $n$ & Blaze avg (s) & Mercury avg (s) & Slowdown\\
\midrule
condition-450-rare        & 391      & 391      & 0.070 & 75.71  & 1{,}082$\times$\\
condition-all             & 99{,}777 & 99{,}777 & 1.543 & 44.73  & 29$\times$\\
condition-ten-frequent    & 96{,}397 & 96{,}397 & 0.121 & 4.70   & 39$\times$\\
condition-ten-rare        & 391      & 391      & 0.060 & 3.10   & 52$\times$\\
condition-two             & 8{,}589  & 8{,}589  & 0.066 & 1.63   & 25$\times$\\
observation-17861-6       & 2{,}515  & 2{,}515  & 0.063 & 1.33   & 21$\times$\\
observation-44261-6       & 35{,}996 & 35{,}996 & 0.087 & 1.71   & 20$\times$\\
observation-72514-3       & 99{,}815 & 99{,}815 & 0.119 & 11.73  & 99$\times$\\
observation-788-0         & 2{,}242  & 2{,}242  & 0.056 & 1.32   & 24$\times$\\
observation-8310-5        & 60{,}092 & 60{,}092 & 0.092 & 1.23   & 13$\times$\\
obs-body-weight-10        & 6{,}701  & 6{,}701  & 0.072 & 7.60   & 106$\times$\\
obs-body-weight-50        & 47{,}953 & 47{,}953 & 0.097 & 7.84   & 81$\times$\\
obs-body-weight-100       & 99{,}807 & 99{,}807 & 0.130 & 8.61   & 66$\times$\\
calcium-date-age          & 18{,}698 & 18{,}698 & 0.096 & 13.56  & 141$\times$\\
hemoglobin-date-age       & 13{,}781 & 13{,}781 & 0.077 & 4.13   & 54$\times$\\
inpatient-stress          & 1{,}628  & 1{,}628  & 1.133 & 110.82 & 98$\times$\\
stratifier-condition-code & 5{,}658{,}095 & 5{,}658{,}095 & 3.063 & 41.98 & 14$\times$\\
\bottomrule
\end{tabular}
\end{table}

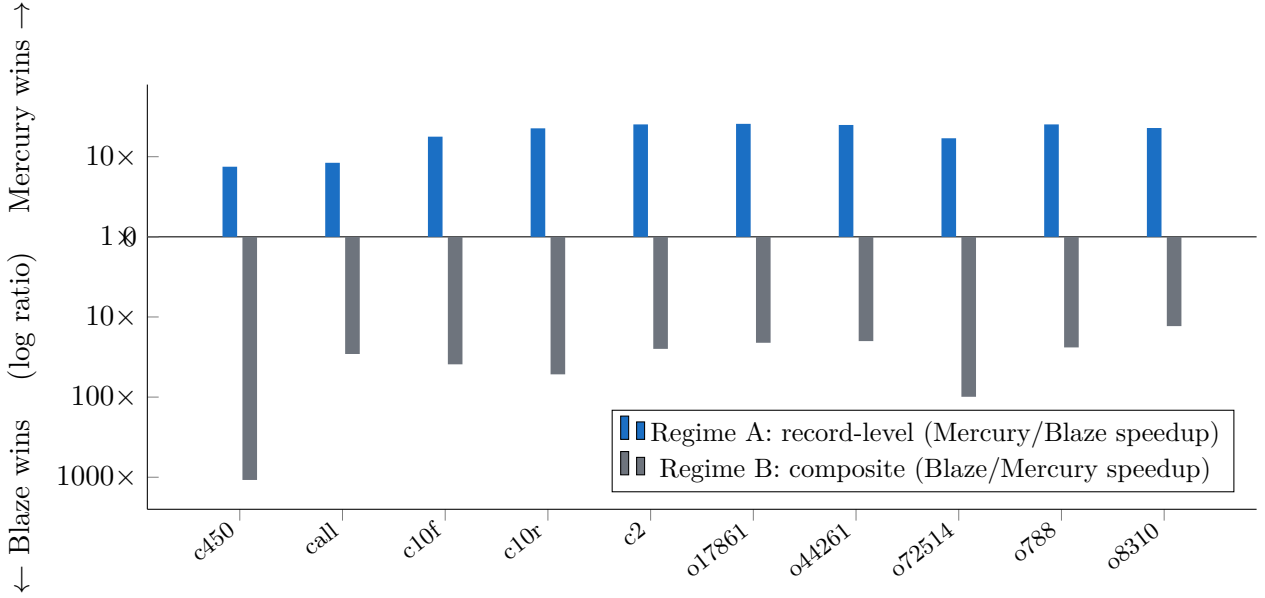
\begin{figure}[H]
\centering
\begin{tikzpicture}
\begin{axis}[
    ybar, bar width=5.5pt,
    width=\linewidth, height=7.2cm,
    ylabel={$\leftarrow$ Blaze wins \quad (log ratio) \quad Mercury wins $\rightarrow$},
    symbolic x coords={c450,call,c10f,c10r,c2,o17861,o44261,o72514,o788,o8310},
    xtick=data, x tick label style={rotate=38,anchor=east,font=\footnotesize},
    ymin=-3.4, ymax=1.9,
    ytick={-3,-2,-1,0,1},
    yticklabels={1000$\times$,100$\times$,10$\times$,1$\times$,10$\times$},
    axis lines*=left,
    legend style={at={(0.98,0.04)},anchor=south east,font=\small},
    every axis plot/.append style={draw=none},
    extra y ticks={0}, extra y tick style={grid=major, grid style={black,thin}},
]
\addplot[fill=mercblue, draw=none] coordinates {
 (c450,0.875) (call,0.924) (c10f,1.250) (c10r,1.354) (c2,1.403)
 (o17861,1.410) (o44261,1.396) (o72514,1.230) (o788,1.403) (o8310,1.358)};
\addplot[fill=blazegray, draw=none] coordinates {
 (c450,-3.034) (call,-1.462) (c10f,-1.591) (c10r,-1.716) (c2,-1.398)
 (o17861,-1.322) (o44261,-1.301) (o72514,-1.996) (o788,-1.380) (o8310,-1.114)};
\legend{Regime A: record-level (Mercury/Blaze speedup), Regime B: composite (Blaze/Mercury speedup)}
\end{axis}
\end{tikzpicture}
\caption{Two regimes, one architecture. For each query, Mercury's record-level speedup is
plotted upward and its composite slowdown downward (signed $\log_{10}$ ratio). The same
storage decision produces both halves; no query escapes the mirror. The downward half
measures a mode Mercury does not even ship enabled --- unscoped composite evaluation is off
by default and reports \emph{not supported}; it was switched on for this test. The deepest
bar (\texttt{c450}, 1{,}082$\times$) is the 450-code retrieve --- the shape Future Work
item~4 targets from both directions at once.}
\label{fig:regimes}
\end{figure}

Figure~\ref{fig:regimes} superimposes both regimes on a single signed log-ratio axis: the
mirror image \emph{is} the finding.

\section{The correctness journey}\label{sec:journey}
The first campaign scored 10/21 exact-count matches. Every failure traced to three root
causes. \textbf{(1) Dropped code filters}: named \texttt{code}/\texttt{concept} definitions
were never bound, and unresolvable valuesets fell through unfiltered, so retrieves matched
everything; counts were wrong \emph{loudly} and count comparison caught them. The fix also
made unresolved terminology a hard evaluation error, matching Blaze's honesty. \textbf{(2)
urn:uuid split-keying at ingest}: Synthea transaction bundles reference subjects as
\texttt{urn:uuid:X}; Mercury keyed clinical rows under the literal string and Patient rows
under the bare id. Record-level evaluation under the bare id found no data --- yet in-engine
composites were \emph{accidentally correct} because the registry enumerated both key forms.
Composite comparison could never catch this; the agreement gate did (59.97\,\% agreement is
an unmistakable alarm). \textbf{(3) Un-rewritten intra-bundle references}: FHIR transaction
rules require \texttt{urn:uuid} references to be rewritten to \texttt{Type/id}; Mercury
stored them literally, so reference-string joins
(\texttt{C.encounter.reference = 'Encounter/' + E.id}) matched nothing and
\texttt{inpatient-stress} returned 0 against Blaze's 1{,}628 --- silently wrong, the worst
class. A related evaluation gap (choice-type \texttt{value[x]} compared against a Quantity
hard-erroring instead of yielding null per ELM cast semantics) was fixed with spec-compliant
null-coercion while keeping genuinely ill-typed comparisons loud.

After fixes and a full re-ingest: 17/17 exact composites and 100.000\,\% record-level
agreement. The \emph{store-hygiene bonus}: repairing the split-key layout made every
per-patient lookup 1.2--2.8$\times$ faster (e.g.\ 2.82\,ms $\to$ 1.04\,ms p50) and the full
fan-out 48\,s $\to$ 41\,s --- the bug had been taxing every read with double-key resolution.

\section{Attribution: why both regimes are true at once}\label{sec:attribution}
Measured on the live 112\,M-resource store, Mercury's single-subject evaluation costs p50
227\,$\mu$s, of which storage reads are $\approx$1\,\% (patient-scoped point-gets,
2--5\,$\mu$s); the remainder is per-subject environment, terminology binding, and AST
interpretation --- repeated once per subject. Blaze compiles once and pays
$\approx$0.63\,$\mu$s per patient amortized on the population axis, but $\approx$27\,ms per
call on the subject axis. One architectural choice, two mirror-image bills.

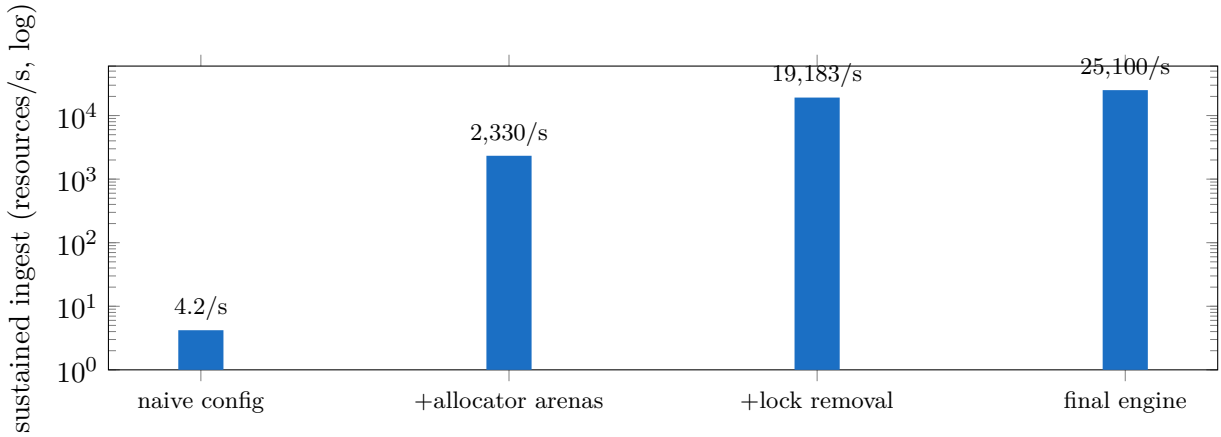
\begin{figure}[H]
\centering
\begin{tikzpicture}
\begin{axis}[
    ybar, bar width=17pt, ymode=log,
    width=\linewidth, height=5.6cm,
    ylabel={sustained ingest (resources/s, log)},
    symbolic x coords={naive config,+allocator arenas,+lock removal,final engine},
    xtick=data, x tick label style={font=\footnotesize},
    ymin=1, ymax=60000,
    nodes near coords, every node near coord/.append style={font=\footnotesize},
    point meta=explicit symbolic,
    every axis plot/.append style={draw=none},
]
\addplot[fill=mercblue, draw=none] coordinates {
 (naive config,4.2)      [4.2/s]
 (+allocator arenas,2330) [2{,}330/s]
 (+lock removal,19183)   [19{,}183/s]
 (final engine,25100)    [25{,}100/s]};
\end{axis}
\end{tikzpicture}
\caption{The ingest journey (log scale): a $\sim$6{,}000$\times$ path from the naive
configuration to the final engine. The largest single factor was the memory allocator's arena
policy, not any database code. Blaze's sustained transaction-bundle rate on the same corpus
is $\approx$6{,}800\,res/s (4\,h\,36\,m end-to-end) --- Mercury's final NDJSON path lands
3.7$\times$ faster end-to-end.}
\label{fig:ingest}
\end{figure}

Ingest attribution on the same corpus (Figure~\ref{fig:ingest}): glibc malloc arena policy
(\texttt{MALLOC\_ARENA\_MAX} 2$\to$64) $\approx$8$\times$; global-lock removal
$\approx$3.5--4$\times$ in the parallel phase but 1.3$\times$ end-to-end (a shared
single-threaded tail --- Amdahl's law applied twice in one experiment); RocksDB
merge-operator vs.\ read-modify-write $\approx$ nil. Final ingest: 112.3\,M resources in
74.7\,min including spec-required reference rewriting (Blaze: 4\,h\,36\,m via transaction
bundles --- different paths, declared as such).

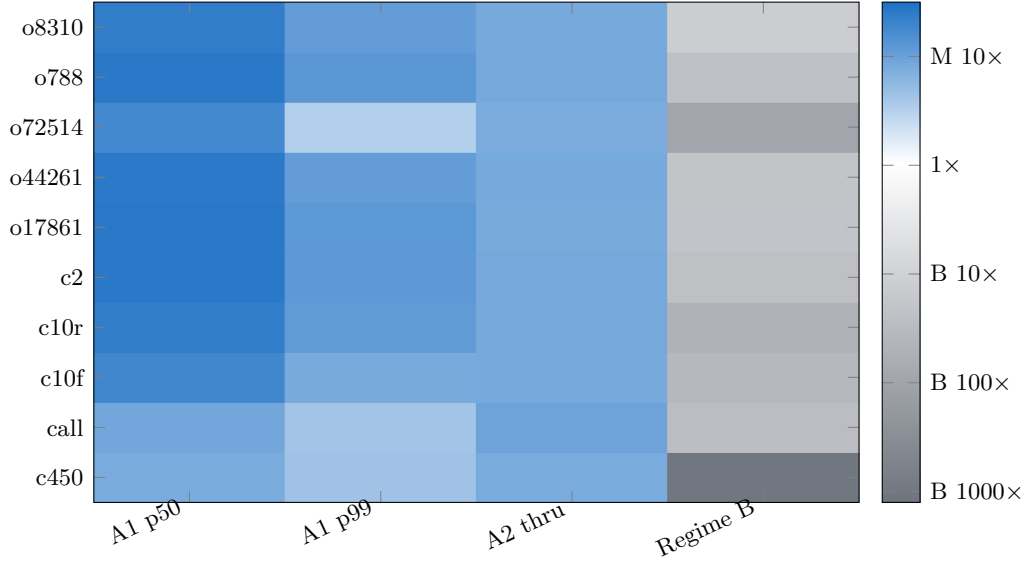
\begin{figure}[H]
\centering
\begin{tikzpicture}
\begin{axis}[
    width=0.72\linewidth, height=8.2cm,
    colormap={div}{color(0cm)=(blazegray); color(3.1cm)=(white); color(4.6cm)=(mercblue)},
    point meta min=-3.1, point meta max=1.5,
    colorbar, colorbar style={
      ytick={-3,-2,-1,0,1},
      yticklabels={B 1000$\times$,B 100$\times$,B 10$\times$,1$\times$,M 10$\times$},
      font=\footnotesize},
    xtick={1,2,3,4},
    xticklabels={A1 p50,A1 p99,A2 thru,Regime B},
    x tick label style={rotate=25,anchor=east,font=\footnotesize},
    ytick={1,...,10},
    yticklabels={c450,call,c10f,c10r,c2,o17861,o44261,o72514,o788,o8310},
    y tick label style={font=\footnotesize},
    enlargelimits=false, axis on top,
]
\addplot[matrix plot*, mesh/cols=4, point meta=explicit] coordinates {
 (1,1)[0.875] (2,1)[0.633] (3,1)[0.875] (4,1)[-3.034]
 (1,2)[0.924] (2,2)[0.613] (3,2)[0.959] (4,2)[-1.462]
 (1,3)[1.250] (2,3)[0.886] (3,3)[0.898] (4,3)[-1.591]
 (1,4)[1.354] (2,4)[1.041] (3,4)[0.902] (4,4)[-1.716]
 (1,5)[1.403] (2,5)[1.061] (3,5)[0.902] (4,5)[-1.398]
 (1,6)[1.410] (2,6)[1.076] (3,6)[0.889] (4,6)[-1.322]
 (1,7)[1.396] (2,7)[1.029] (3,7)[0.900] (4,7)[-1.301]
 (1,8)[1.230] (2,8)[0.505] (3,8)[0.865] (4,8)[-1.996]
 (1,9)[1.403] (2,9)[1.083] (3,9)[0.908] (4,9)[-1.380]
 (1,10)[1.358](2,10)[1.029] (3,10)[0.891] (4,10)[-1.114]
};
\end{axis}
\end{tikzpicture}
\caption{Every measured ratio in one view (signed $\log_{10}$; blue = Mercury faster, gray =
Blaze faster). The regime boundary is a vertical line: three record-level metrics are
uniformly blue, the composite column uniformly gray. No query crosses it --- the tradeoff is
architectural, not query-specific.}
\label{fig:heat}
\end{figure}

Figure~\ref{fig:heat} compresses every measured ratio into one view of the regime boundary.

\section{Discussion}\label{sec:discussion}
\paragraph{Regime-match, not engine-superiority.} Decision support and atom-level
measurement (millisecond budgets, inspectable individual determinations) fit record-level
engines; whole-store composite reporting (minute budgets, no data movement) fits compiled
scan engines. The procurement question is ``which regime am I buying for?'' --- and, within
quality measurement, ``do I need the atoms or only the score?''

\paragraph{Methodology.} Run the defender's own suite under the defender's protocol; add
record-level agreement gates (two of our three bug classes were invisible to composite
comparison); publish the failure taxonomy with the same prominence as the speedups.

\paragraph{Threats to validity.} Single corpus generator (Synthea); cohort-scoring boolean
measures only; Blaze measured once on its documented configuration without further tuning
(its published numbers are consistent with ours); Blaze's \texttt{\$cql} endpoint (beta) was
unavailable in this deployment, so the CDS-surface comparison rests on
\texttt{\$evaluate-measure}; engines measured serially on the same box class, not
simultaneously; one Mercury latency-tail note (a single 59\,ms call among 2{,}000 on
\texttt{observation-72514-3}, p95 3.8\,ms).

\section{Future work}\label{sec:future}
In-engine composite aggregation is not what Mercury is for --- but we thought it would be fun
to find out, and it shows: Blaze does an excellent job in that regime. The gap is closable
without abandoning the patient-first layout, in escalating order of ambition: (1)~hoist
per-subject setup (estimated 227\,$\mu$s $\to$ 20--50\,$\mu$s per subject);
(2)~composite-shape recognition in the planner, answering boolean code-exists forms from
index postings ($\approx$100\,k point-gets $\approx$ 10--50\,ms parallel --- competitive with
Blaze's 63\,ms on the 2.5\,\% query); (3)~a value-first auxiliary index
(\texttt{type|property|value}$\to$patients) making population enumeration a single prefix
scan, backfillable through the existing reconcile machinery; (4)~multi-code retrieves via
ordered prefix scan (removing the 450-point-get shape behind both the 1{,}082$\times$
composite outlier and the 9.7\,ms Regime-A entry); (5)~stratifier streaming (removing the
disclosed DNF); (6)~\emph{streamed result output}: large evaluations currently pay full
materialization and serialization of the report before the first byte returns --- the cost
behind both the subject-list report shape and the stratifier DNF. A streaming mode would
return the aggregate immediately and then serialize matches and per-stratum intermediates
incrementally, so an analyst can obtain the answer \emph{and} re-run the same query streaming
out its intermediate matches for inspection --- without the engine ever holding the full
result set in memory. This composes with~(5), and it is the natural API for an engine whose
thesis is that the atoms matter: intermediate results become an analysis feature rather than
a memory liability. Beyond the engine: CQF Ruler as a third subject, CV/ratio measures,
non-Synthea corpora, and \emph{CQF-Bench} --- a cross-engine benchmark of the full CQF
record-level operation surface (Section~\ref{sec:background}) that
\texttt{\$evaluate-measure}-centric suites, this one included, leave unmeasured.

\section{Conclusion}\label{sec:conclusion}
Two engines, one corpus, one machine class, and the defender's own benchmark suite --- and
each engine wins by one to three orders of magnitude in the regime it was built for, with
record-level answers agreeing 10{,}000/10{,}000. The tradeoff is architectural, predictable,
and measurable --- provided benchmarks test both regimes and hold correctness to a
record-level standard.

\paragraph{Data availability.} Engine and harness: \texttt{carreraGroup/mercury}
(\texttt{scripts/blaze-native-bench/}). Query suite and protocol: \texttt{samply/blaze}
\texttt{docs/performance}, used verbatim. Result artifacts (9-run raw durations, per-patient
booleans for both engines, identifier map, sample checksums, store snapshots, full pipeline
logs) are archived in S3; release mode (public bucket vs.\ on-request) to be finalized with
the preprint.

\end{document}